\documentclass[11pt,fleqn]{elsart}

\usepackage{graphicx,psfrag,here,epsfig}
\usepackage{amsmath,amssymb}
\usepackage{exscale}
\usepackage{longtable,multirow}

\allowdisplaybreaks[1]

\newfont{\hermes}{cmtt10 at 10pt}

\newcommand{\kreis}[1]{\unitlength1ex\begin{picture}(2.5,2.5)%
\put(0.75,0.75){\circle{2.5}}\put(0.75,0.75){\makebox(0,0){#1}}\end{picture}}

\newcommand{\bea}{\begin{eqnarray}}
\newcommand{\eea}{\end{eqnarray}}

\newcommand{\be}{\begin{equation}}
\newcommand{\ee}{\end{equation}}
\newcommand{\een}{\nonumber\end{equation}}
\newcommand{\ed}{\end{document}}

\newcommand{\scs}{\; ,\;}
\newcommand{\fs}{\; .}

\renewcommand{\theequation}{\arabic{equation}}

\begin{document}

\renewcommand{\theequation}{\arabic{equation}}
\begin{frontmatter}

\hfill{Accepted for publication in Phys. Lett.  B}

\title{\Large\bf On the pion decay constant }

{J.~Gasser}\footnote{Corresponding author.}\,\,and\,\,
{G.R.S.~Zarnauskas}\footnote{\,\,Present address:
Instituto de F\'{\i}sica, Universidade de S\~ao Paulo,
C.P. 66318, 05315-970, S\~{a}o Paulo, SP, Brazil.\\[1mm]
E-mail addresses: gasser@itp.unibe.ch; gabrielz@if.usp.br}\\[2mm]
{Albert Einstein Center for Fundamental Physics,
\\Institute for Theoretical Physics, University of Bern,\\
Sidlerstr. 5, 3012 Bern, Switzerland}

\begin{abstract}
The pion decay constant $f_\pi$ plays a crucial role in many areas of low energy particle physics.
 Its value may e.g. be deduced {}from experimental data on leptonic pion decays.
Here, we provide comments on several aspects of this evaluation. In particular, we point out that at the 
present level of experimental accuracy, the value of $f_\pi$  is sensitive to the value 
of the pion mass chosen in its chiral expansion.
\end{abstract}
\setcounter{footnote}{0}

\end{frontmatter}


\section{Introduction}
\vspace{-2\parskip}
In the framework of  QCD, the pion decay constant $f_\pi$ may be defined
through the coupling of the axial current to the pion~\cite{PDG},
\be\label{eq:fpiqcd}
\langle 0|A_\mu(0)|\pi^-(p)\rangle=ip_\mu f_\pi\,;\,
 A_\mu=\bar u \gamma_\mu\gamma_5 d\scs
\ee
where $|\pi^-(p)\rangle$ denotes a one-pion state with normalization  $\langle \pi^-(p')|\pi^-(p)\rangle=2(2\pi)^3p^0\,
\delta^{(3)}(\mathbf{p}'-\mathbf{p})$.

The pion decay constant  plays a crucial role in many areas of low energy 
particle physics. First of all, it dictates the strength of leptonic pion decays,
\be\label{eq:nonradiative}
\pi^-\to \ell^-\bar\nu_\ell\,;\,\ell=e,\mu\scs
\ee
with rate\footnote{To ease notation, we often write in the following $\pi\to\ell\nu$, which stands for $\pi^-\to\ell^-\bar\nu_\ell$ or  for $\pi^+\to\ell^+\nu_\ell$.} 
\be
\Gamma^{(0)}(\pi\to \ell\nu)=\frac{G_F^2|V_{ud}|^2f_\pi^2}{8\pi} m_\pi m_\ell^2\left(1-\frac{m_\ell^2}{m_\pi^2}\right)^2
\ee
in the absence of electromagnetic interactions.
 It also enters    the Goldberger-Treiman relation~\cite{goldbergertreiman},
\begin{equation}
f_\pi\, g_{\pi\! N}=
\sqrt{2}m_p \,g_A\scs
\end{equation}
which relates the weak and strong  coupling constants $f_\pi,g_A$  and $g_{\pi\! N}$ 
 with the proton mass $m_p$. This relation is exact in the chiral limit $m_u=m_d=0$~\cite{nambu60} and
  may  well be considered the starting point of precise low energy 
hadron physics. It has been 
and still is a crucial test of our understanding of low energy hadron dynamics. Further, $f_\pi$ happens to 
determine the strength of $\pi\pi$ interactions and thus acts as an (inverse) expansion parameter in Chiral Perturbation Theory (ChPT) ~\cite{effective1,effective2}.
Last but not least, $f_\pi$ is now amenable  to lattice calculations, 
see e.g. Ref.~\cite{TMLQCD} for an evaluation with $N_f=2+1+1$, and for further 
references.

The most precise determination of $f_\pi$ is presently obtained {}from
leptonic decays of the pion. In fact,
experiments have reached a level of precision which makes it mandatory to include 
radiative corrections, and to consider  the rate for
\be\label{eq:radiative}
\pi\to \ell\nu(\gamma)\fs
\ee
According to PDG~\cite{PDG}, 
\be\label{eq:valueoffpi}
f_\pi=\left(130.4\pm 0.04\pm 0.2\right)\,\mbox{MeV}\fs
\ee
This value is  based  on data [branching fraction for $\pi^-\to\mu^-\bar\nu_\mu (\gamma)$ and lifetime], 
and on  theoretical work performed in Refs. 
\cite{marciano,urech,neufeld96,bijnensprades,moussallam,knecht,ananthanarayanmoussallam,descotesgenon,cirigliano}.
The first (second) uncertainty is due to the uncertainty in the value of $V_{ud}$ (to
the  uncertainties in the  higher order corrections in the evaluation of 
the matrix element for the decay (\ref{eq:radiative})).
It is the main purpose of this Letter  to   comment on recent calculations~\cite{marciano,neufeld96,knecht,cirigliano} of the transition 
matrix element for this process, and on the value of $f_\pi$ 
reported in Eq.~(\ref{eq:valueoffpi}), see Section~\ref{sec:questions} for 
details on the  questions investigated here.

\section{The effective Lagrangian}
\vspace{-2\parskip}
A very elegant and convenient tool to perform the 
calculation is the effective field theory framework set up in 
Refs.~\cite{urech,neufeld96,knecht,cirigliano}.
We adhere here to this method, and 
come back to its relation to the underlying theory below.  
 The pertinent lowest-order effective  Lagrangian for three flavours can be found  in
Ref.~\cite{knecht}. Here, we consider its two-flavour version,
\begin{align}\label{eq:Leff_leading}
{\mathcal L}_{eff} =& \frac{F^2}{4}
\left\langle u_\mu u^\mu+\chi_+ \right\rangle +
e^2F^4Z \left\langle u^\dagger Q u^2Qu^\dagger \right\rangle
\nonumber\\
&-\frac{1}{4}F_{\mu\nu}F^{\mu\nu}-\frac{1}{2\xi}(\partial_\mu A^\mu)^2+\sum_{\ell} \left[ \bar{\ell} \left(
i\slash\!\!\!\partial
+ e\slash\!\!\!\!A - m_\ell \right) \ell+ \bar{\nu}_{\ell L} i
\slash\!\!\!\partial \nu_{\ell L} \right] \ , 
\end{align}

\noindent where the flavour-trace is indicated by $\langle\rangle$, and
\begin{align}
u=e^{i \phi/2F} \ \ \ , \ \ \ \
\phi= \begin{pmatrix}
\pi^0 & \sqrt{2} \pi^+  \\
\sqrt{2} \pi^- & -\pi^0
\end{pmatrix} \ .
\end{align}
Here, $F$ denotes the pion decay constant 
in the chiral limit, in a normalization which is standard in ChPT:  Let $f={f_{\pi}}_{|{m_u=m_d=0}}$. Then $f=\sqrt{2}F$.  In the absence of pseudoscalar densities, one has
\begin{align}
\chi_+=u^\dagger\chi u^\dagger+u\chi u \ \ \ , \ \ \ \ \chi=2B\,
\mbox{diag}(m_u,\ m_d) \ .
\end{align}
\noindent 
The external vector and axial vector external sources $v_\mu$ and $a_\mu$ 
contain also the lepton and photon fields,
\begin{align}
u_\mu&=i\left[u^\dagger\left( \partial_\mu-ir_\mu \right)u - u\left(
\partial_\mu-i l_\mu \right) u^\dagger \right]\ ,\nonumber\\
l_\mu & =v_\mu-a_\mu-eQA_\mu+\sum_{\ell = e,\mu} \left( \bar{\ell} \gamma_\mu
\nu_{\ell L} Q^W+h.c.\right)\ ,\nonumber\\
r_\mu & =v_\mu+a_\mu-eQA_\mu \ ,
\end{align}

\noindent with
\begin{align}
Q=\frac{1}{3}
\begin{pmatrix}
2 & 0  \\
0 & -1
\end{pmatrix} \ \ \ , \ \ \ \
Q^W=-2\sqrt{2}G_F\begin{pmatrix}
0 & V_{ud}  \\
0 & 0
\end{pmatrix} \ .
\end{align}
For further notation, and for the terms at next-to-leading order, 
see Ref.~\cite{knecht}.
\section{$\pi\to \ell\nu$ without electromagnetic interactions}
\vspace{-2\parskip}
We first consider the non-radiative decay $\pi\to \ell\nu$ in the absence of electromagnetic corrections, and set  $e=0$ in the effective Lagrangian. 
It is convenient to use the axial current as an interpolating field for the pion, 
\be\label{eq:LSZ}
\langle \bar\nu_\ell(q_2)\ell^-(q_1);\mbox{out}|A_\mu(0)^\dagger|0\rangle 
=\frac{-i f_\pi p_\mu}{m_\pi^2-p^2} 
T(q_1,q_2)\, ; \, p=q_1+q_2\fs
\ee
At $p^2=m_\pi^2$, the quantity $T(q_1,q_2)$ is the transition 
amplitude for $\pi^-\to \ell^-\bar\nu_\ell$, 
\be\label{eq:transitionampl}
T=i{\sqrt{2}G_F}V_{ud}f_\pi m_\ell\bar u(q_1)v_L(q_2)\fs
\ee
As has been pointed out by the authors of Ref.\cite{cirigliano}, the chiral 
expansion of the transition amplitude, evaluated in the effective field theory 
framework mentioned,
amounts to the chiral expansion of $f_\pi$. A proof of the statement is 
provided in the Appendix. It relies on the fact that the pion decay constant 
also occurs in the correlator of two axial currents,
\vskip-5mm
\bea\label{eq:twopoint0}
A^{\mu\nu}&= &i\int d^4x e^{ipx}\langle 0|TA^\mu(x) A^{\nu\dagger}(0)|0\rangle
=B(p^2)p^\mu p^\nu + C(p^2)g^{\mu\nu}\scs\\
\label{eq:twopoint}
B(p^2)&=&\frac{f_\pi^2}{m_\pi^2-p^2}+R(p^2)\fs
\eea
\vskip-5mm
The quantity $R$ is holomorphic in the complex $p^2$-plane, cut along the real axis for $p^2\geq 9m_\pi^2$. [The ambiguities inherent in the definition of $A^{\mu\nu}$, 
generated by the short distance singularities in $\langle 0|T A^\mu(x) A^{\nu\dagger}(0)|0\rangle$, 
affect $R$ and $C$ only.] 
One could as well use the relation (\ref{eq:twopoint}) to define  $f_\pi$ in pure QCD.

\section{Two questions}\label{sec:questions}
\vspace{-2\parskip}
Because the effects of real and virtual photons have to be included for the confrontation of theory with 
experiment, infrared (IR) singularities occur in intermediate steps 
of the calculation. One source of these  singularities is the fact
 that the correlator $A^{\mu\nu}$, evaluated 
 at $e\neq 0$, develops a branch point 
in the form factor $B(p^2)$ at $p^2=m_\pi^2$, as a result of which
 there is no isolated pole-contribution -- the decomposition Eq.~(\ref{eq:twopoint}) 
does not hold anymore in the presence of electromagnetic interactions. In addition,
the formula Eq.~(\ref{eq:LSZ}) cannot be used without further ado when $e\neq 0$.
So, one may wonder about the role of  the pion decay constant 
that is determined in pion decays: 
\begin{enumerate}
\item[i)]
 What is its relation to the correlator $A^{\mu\nu}$ at $e\neq 0$?
\item[ii)]
What is  its relation 
to the pion decay constant $f_\pi$ in pure QCD, as it occurs in the decomposition Eq.~(\ref{eq:twopoint})?
\end{enumerate}
 To the best of our knowledge,
 these two questions  were never discussed in  full detail in the literature, 
and we find it instructive to shed additional light on the issue. 
 On the other hand, 
we do not   question the final algebraic result for the rate as provided in the works 
mentioned -- we have nothing to add here.

\section{Photons generate a branch point}
\vspace{-2\parskip}
\subsection{Infrared regularization}
\vspace{-2\parskip}
To perform the  calculations, one may tame
the IR singularities by providing the photon with a small mass $m_\gamma$, such that
the decomposition Eq.~(\ref{eq:twopoint}) still holds, with
$f_\pi\to \bar f_\pi$, where the constant $\bar f_\pi$ now also includes contributions {}from virtual photons.
  The remainder $R(p^2)$  generates a branch point at $p^2=(m_\pi+m_\gamma)^2$. 
The quantity  $\bar f_\pi$  is gauge dependent and 
diverges logarithmically as the photon mass is sent to zero.
 Further, the LSZ formula Eq.~(\ref{eq:LSZ}) 
remains true, with $f_\pi\to\bar f_\pi$.
The IR singularities cancel at the end when adding the rates 
for $\pi\to \ell\nu$ and for $\pi\to \ell\nu\,(n\gamma)$, 
and sending  $m_\gamma$ to zero at the very end of the calculation provides 
the desired result. This is the method used in Refs.~\cite{neufeld96,knecht,cirigliano}. 

Because dimensional regularization is a very useful ultraviolet and infrared regulator 
for our  purpose, we adhere in the following to this alternative 
regularization~\cite{'t hooft,bollini,rad1,rad2}, where the photon mass is set to zero 
{}from the very beginning. We start the discussion with the 
evaluation of the correlator
\be\label{eq:twopoint_e}
A_e^{\mu\nu} = i\int d^d\!x e^{ipx}\langle 0|TA^\mu(x) A^{\nu\dagger}(0)|0\rangle_e
\ee
in $d$ space-time dimensions, in the presence of electromagnetic interactions, in the framework of the effective theory 
defined by the Lagrangian Eq.~(\ref{eq:Leff_leading}). The index $e$ indicates that virtual photons are  included.
\begin{figure}
{\epsfig{file=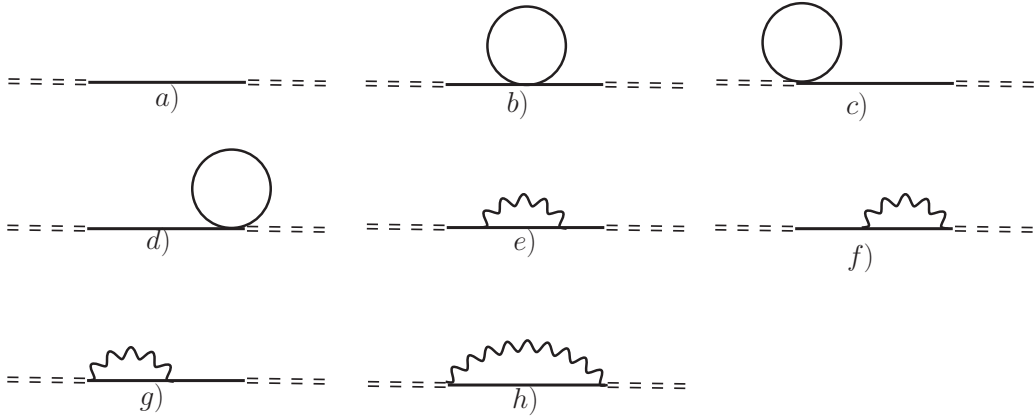,width=\textwidth}}
\caption{Diagrams that contribute to the correlator $A_e^{\mu\nu}$ at one-loop order. Double lines denote axial currents, solid (wavy) lines  pions (photons). Contributions 
{}from counterterms are not shown, nor do we indicate graphs which generate 
local terms only.}\label{fig:twopoint}
\end{figure}

\subsection{Loop contributions at $d\neq 4$: explicit expressions}
\vspace{-2\parskip}
To keep everything as simple as possible,  we restrict ourself to a 
one-loop calculation. Some of the pertinent graphs which contribute to $A_e^{\mu\nu}$ are displayed in Fig.~\ref{fig:twopoint}. 
 A typical contribution generated e.g. by diagrams Figs.~\ref{fig:twopoint}a+\ref{fig:twopoint}e reads
\be\label{eq:amunu_example}
 A_e^{\mu\nu}
=\frac{2p_\mu p_\nu F^2}{M^2-p^2}\left(1+\frac{e^2M^2J(p^2)}{M^2-p^2}\right)+\cdots\scs
\ee
 where $J(p^2)$ denotes the one-loop integral
\be\label{eq:oneloopintegralJ}
J(p^2)=\frac{1}{i}\int \frac{d^dl}{(2\pi)^d}\frac{1}{(M^2-(p-l)^2)(-l^2)}\fs
\ee
 As for the notation, we recall that the chiral expansion of the charged
pion mass starts out with  $M_{\pi^+}^2=M^2+O(e^2,m_q^2)$, 
where $M^2=(m_u+m_d)B$, see Refs.~\cite{urech,neufeld96}.
 The ellipsis in Eq.~(\ref{eq:amunu_example}) stands for  additional contributions.
 The double pole at $p^2=M^2$  is removed 
by mass renormalization in the standard manner.
We  split off the part which contributes to the pion mass, 
\be
J(p^2)=J(M^2)+(1-z)\underline J(p^2)\,;\, z=\frac{p^2}{M^2}\scs
\ee
such that
\be
 A_e^{\mu\nu}
=\frac{2p_\mu p_\nu F^2}{M_1^2-p^2}(1+e^2\underline J(p^2))+\cdots\,;\, M_1^2=M^2(1-e^2J(M^2))\fs
\ee
If the photon were massive, the photon propagator in Eq.~(\ref{eq:oneloopintegralJ}) would be 
replaced by $1/(-l^2)\to 1/(m_\gamma^2-l^2)$, and the corresponding loop function $\underline J(p^2,m_\gamma)$ would be of the form
$\underline J(p^2,m_\gamma)=C+O(p^2-M^2)$, with $C$ a finite constant at $d=4$, and the decomposition  Eq.~(\ref{eq:twopoint}) would hold  for $A_e^{\mu\nu}$ as well. 
 However, massless photons render loop contributions singular at threshold.
To investigate the structure of $A_e^{\mu\nu}$, 
we evaluate  loop integrals  at $d\neq 4, p^2 \neq  M^2$. 
In particular, the quantity   $J(p^2)$ can be expressed in terms of hypergeometric 
functions \cite{gegeliaetal}.  Expanding the parts which are regular at $p^2=M^2$,  
we find 
\begin{align}
&J(p^2)=
M^{2w}f(w)\left\{1-\frac{\Gamma(-2w)}{\Gamma(-w)}(1-z)\left[(1-z)^{2w}g_1(w,z)+g_2(w,z)\right]\right\}\scs\nonumber\\
&f(w)=\frac{\Gamma(-w)}{1+2w}\frac{1}{(4\pi)^{2+w}}\scs g_1=\frac{\Gamma(1+w)}{z^{1+w}}\scs\nonumber\\
& g_2=-\sum_{n\geq 1}\frac{\Gamma(n-w)}
{\Gamma(n-2w)}(1-z)^{n-1}=-g_1+w^2h(d,z)\scs w=\frac{d}{2}-2\fs
\end{align}
The function $h(d,z)$ is regular at $d=4,z=1$, and  $\underline J(p^2)$ behaves as
\begin{figure}\begin{center}
{\epsfig{file=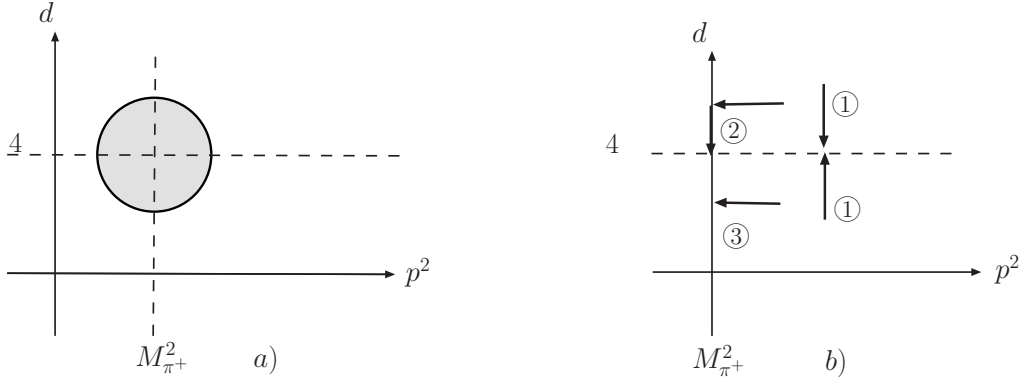,width=\textwidth}}
\caption{The one-loop integral $\underline J$ and the 
correlator $A_e^{\mu\nu}$ in the $d,p^2$ plane. In Fig.~a), the 
shaded circle indicates the region where the expression Eq.~(\ref{eq:barJ}) is 
valid. In Fig.~b), we indicate 3 different limiting procedures. 
The limit  1 (2)  is relevant for renormalization 
 (for the LSZ formalism). 
In the limit 3, $\underline J$ and $A_e^{\mu\nu}$ do not exist. For $\underline J$ use $M_{\pi^+}^2=M^2$.}\label{fig:limits}
\end{center}
\end{figure}
\be
\begin{split}\label{eq:barJ}
\underline J(p^2)&=(L+k)\left[(1-z)^{d-4}-1\right]z^{-1}+\cdots\scs\\
L&=\frac{\mu^{d-4}}{N}\left[\frac{1}{d-4}-\frac{1}{2}\left(\ln{4\pi}+\Gamma'(1)+1\right)\right]\scs 
k=\frac{1}{2N}\left[\ln{\frac{M^2}{\mu^2}}-1\right]\scs\\
N&=16\pi^2
\end{split}
\ee
in the vicinity of $p^2=M^2, d=4$.
The renormalization scale is denoted by $\mu$, and the ellipsis stands for terms that are irrelevant in the following.  The region where this representation is valid is indicated  in Fig.~\ref{fig:limits}a with the shaded circle [use $M_{\pi^+}^2=M^2$]. 

\subsection{Three limiting procedures}
\vspace{-2\parskip}
We now consider three different limiting procedures~\cite{schweizer}. 
First, we let $d\to 4$ 
off the mass shell $p^2=M^2$. 
This limit is relevant for renormalization  and is indicated by the path\,\, 
\kreis{1} in Fig.~\ref{fig:limits}b. 
Second, for the LSZ formalism, one goes to the mass shell first and then considers $d\to 4^+$ [path\,\, \kreis{2} in the figure], see Refs.~\cite{rad1,rad2,brown}. 
The result is
\be\label{eq:3limitsJ}
\underline J(p^2)=
\left\{\begin{array}{ll}(16\pi^2z)^{-1}\ln{(1-z)}& \,  [\mbox{limit}\hspace{2mm} \kreis{1}]\\
-(L+k)& \,[\mbox{limit}\hspace{2mm}  \kreis{2}]\fs\end{array}\right.
\ee
In the limit \kreis{3}, $\underline J$ does not exist.

 As a result of Eq.~(\ref{eq:3limitsJ}), the correlator 
generates a branch point at $p^2=M_1^2$, 
\be
A_e^{\mu\nu}=
\frac{2p_\mu p_\nu F^2}{M_1^2(1-\frac{p^2}{M_1^2})^{1-\frac{e^2}{16\pi^2} }}
+\cdots [\mbox{limit}\,\,\kreis{1}]\fs
\ee
On the other hand, going on the mass shell $p^2=M_1^2$ at $d>4$ results in
\be 
A_e^{\mu\nu}\,\,=\,\,
\frac{2p_\mu p_\nu F^2}{M_1^2-p^2}(1-{e^2}(L+k)+\cdots)\,\,[\mbox{limit}\,\,\kreis{2}]\fs
\ee
In other words, the standard renormalized contribution generates a branch point in the propagator. Going to the mass shell at $d>4$ results in the standard pole behaviour of the propagator. 
\section{The correlator $A_e^{\mu\nu}$ at one loop}
\vspace{-2\parskip}
Including all the graphs at one-loop order results in the following expression for the  correlator in the first limit,

\vskip-8mm

\bea\label{eq:A_limit1}
A_e^{\mu\nu}&&=
\frac{p^\mu p^\nu \kappa_1^2}{M_{\pi^+}^2(1-\frac{p^2}{M_{\pi^+}^2})^{1+e^2g_1}}+\cdots\,\,\,[\mbox{limit}\,\,\,\,\kreis{1}]\scs\nonumber\\[2mm]
\kappa_1^2&=&2F^2\bigl(1+a_1+e^2b_1+
{\mathcal{O}}[M^4,e^4,M^2e^2]\bigr)\scs\nonumber\\
a_1&=&-\frac{1}{NF^2}\bigl(M^2\ln{\frac{M^2}{\mu^2}}
+M_{\pi^+}^2\ln{\frac{M_{\pi^+}^2}{\mu^2}}\bigr)
+\frac{2M^2}{F^2}l_4^r\scs\nonumber\\
b_1&=&\frac{1}{N}[-3+\xi(1+\ln{\frac{M^2}{\mu^2}})]
+K^r(\mu)\scs\nonumber\\
g_1&=&\frac{1}{N}(6-2\xi)
\scs K^r(\mu)=\frac{20}{9}(k_1^r+k_2^r)+4k_9^r\fs
\eea
Here, $l_4^r \,\,(k_i^r)$ are LECs in the effective Lagrangian at order $p^4$ ($e^2p^2$),
 see Refs.~\cite{effective2} (\cite{meissnermullersteininger,knechturech}). We have used the result Ref.~\cite[Eq.~(C.10)]{gasscirus} for the renormalization 
of $k_9$ in any gauge. 
 The above expression 
 reveals the branch point at $p^2=M_{\pi^+}^2$, with strengths $g_1,\kappa_1$ which 
are gauge dependent. The ellipsis indicates terms that are less singular at 
$p^2=M_{\pi^+}^2$, and terms proportional to $g^{\mu\nu}$. The branch point 
is generated by Fig.~\ref{fig:twopoint}e alone -- this explains the fact that a branch point with 
identical strength $g_1$ is also present in the electron propagator in the framework of 
QED~\cite{brown}, or in the 
two-point function of two charged pion fields in scalar QED~\cite{urech}. In Ref.~\cite{brown}, it is shown that for the electron propagator, 
the leading infrared singularity at any order in $e$ is given  by the coefficient $g_1$.

Performing the second limit gives

\vskip-8mm

\bea 
A_e^{\mu\nu}\,\,&&=\,\,
\frac{p^\mu p^\nu \kappa^2}{M_{\pi^+}^2-p^2}+\cdots\,\,\,[\mbox{limit}\,\,\,\,\kreis{2}]
\scs\nonumber\\
\kappa^2&&=2F^2\bigl(1+a_1+e^2b+{\mathcal{O}}[M^4,e^4,M^2e^2]\bigr)\scs\nonumber\\
b&&=(6-2\xi)L+\frac{1}{N}(3\ln{\frac{M^2}{\mu^2}}-4)+K^r(\mu)\fs
\eea

\vskip-8mm

The ellipsis has the same meaning as before. It is seen that in this case, the 
correlator does have -- aside {}from less singular terms -- 
a pole contribution at $p^2=M_{\pi^+}^2$, with a residue that is 
divergent at $d=4$, and gauge dependent. At $e=0$, the quantity $\kappa=\kappa_1$ coincides with the pion decay constant $f_\pi$, evaluated at one-loop order 
in ChPT~\cite{effective2},
\be\label{eq:fpiwhatmass?}
 f_\pi=\sqrt{2}F\left(1-\frac{M^2}{16\pi^2 F^2}\ln{\frac{M^2}{\mu^2}}
+\frac{M^2}{F^2}l_4^r+{\mathcal{O}}(M^4)\right)\fs
\ee
 
Neufeld and Rupertsberger~\cite{neufeld96}  have evaluated the pion decay constant with the photon mass $m_\gamma$ as an infrared regulator. With the identification
\bea
L\to -\frac{1}{2N}(\ln{\frac{m_\gamma^2}{\mu^2}}+1)\scs
\eea
the quantity $\kappa/\sqrt{2}$, evaluated at $\xi=1$,  agrees with their  $F_{\pi^\pm}$,  
translated to the $SU(2)\times SU(2)$ case by use of the matching relations for the LECs 
$l^r_4,k^r_{1,2,9}$ worked out in Refs.~\cite{effective3,matchinglecs}.

In Refs.~\cite{moussallam,pinzke}, it is shown that the LEC $k_9^r$ depends logarithmically on the scale of the underlying theory [we use the matching of $k_9^r$ to $K_{12}^r$ 
as worked out in Ref.~\cite{matchinglecs}], see also Ref.~\cite{descotesgenon}. In other words, 
the residues $\kappa^2,\kappa_1^2$ are not uniquely defined in the 
framework of QCD+QED.  An analogous scale dependence 
of $k_9^r$ occurs in the effective theory of the linear sigma model, 
coupled to electromagnetism~\cite{gasscirus}.
\section{Width for $\pi\to\ell\nu(\gamma)$}
\begin{figure}\begin{center}
{\epsfig{file=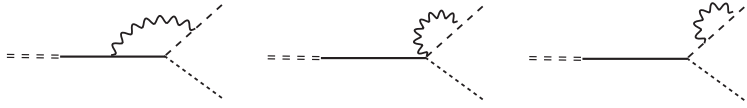,width=10cm}}
\caption{Graphs which contribute to $T_e$ and which contain a virtual lepton. Double (solid) lines: axial current (charged pion). Long  dashed (short dashed) lines: charged 
(neutral) lepton. Wavy lines: photon.}
\label{fig:virtuallepton}
\end{center}\end{figure}

\vspace{-2\parskip}
 We come back to the evaluation of the width for leptonic pion decays                    
(\ref{eq:radiative}). The reduction formula  reads
\be\label{eq:LSZe}
\langle \bar\nu_\ell(q_2)\ell^-(q_1);\mbox{out}|A_\mu(0)^\dagger|0\rangle_e 
=\frac{-i \kappa p_\mu}{M_{\pi^+}^2-p^2} 
T_e(q_1,q_2)\, ; \, p=q_1+q_2
\ee
for the non-radiative part, and similarly for the case when a photon is emitted in addition. Here, it is understood that the limit $p^2\to M_{\pi^+}^2$ is taken at $d>4$.
Let us denote by class I (class II) the set of graphs which do not (which do) contain a virtual charged lepton, and disregard the leptonic counter term contributions for a moment. 
The graphs in class I are obtained by replacing one of the axial currents in 
Fig.~\ref{fig:twopoint} by  $\ell^-\gamma_\mu\bar\nu_{\ell L}$.
{}From the previous discussion and {}from Eq.~(\ref{eq:transitionampl}), it is clear that the graphs in  class I generate the amplitude
\be\label{eq:classCI}
i{\sqrt{2}G_F}V_{ud}\kappa m_\ell\bar u(q_1)v_L(q_2)\equiv C_{I}\kappa\scs
\ee
see also the Appendix.
 The graphs {}from class II are displayed in Fig.~\ref{fig:virtuallepton}. 
These contributions are of 
order $F$ in this order of the momentum expansion. This remains true including  the leptonic counterterms, and 
we denote the sum by  $C_{II}e^2F$.
Finally, one has to add the effect {}from real photon emission -- the pertinent amplitude is denoted by 
$C_{III}eF$. 
All in all, the width at one loop is obtained {}from 
\be\label{eq:everything}
\Gamma(\pi\to \ell\nu\,(\gamma))=\langle\!\langle |C_I\kappa +C_{II}e^2F|^2\rangle\!\rangle 
+\langle\!\langle |C_{III}eF|^2\rangle\!\rangle\scs
\ee
where the symbol $\langle\!\langle\,\rangle\!\rangle$  denotes phase space integrations, including all kinematic factors. 
The IR singularities and the gauge dependence cancel out in  $\Gamma(\pi\to \ell\nu\,(\gamma))$, and one ends up with the expression first given in Eq. (5.1) of Ref.~\cite{knecht}, matched to the $SU(2)\times SU(2)$ case considered here. Further, as pointed out in Ref.~\cite{descotesgenon}, the above mentioned scale dependence of $k_9^r$ is cancelled by the scale dependence of the leptonic LEC $X_6^r$ introduced in Ref.~\cite{knecht}.  

\section{The answers to the questions in Section 
\ref{sec:questions}}
\vspace{-2\parskip}
We are  now prepared to answer the questions raised   in Section \ref{sec:questions}.

{\bf{Answer to question i)}}: 

The relation of the pion decay constant to the correlator of two axial currents is 
evident {}from Eq.~(\ref{eq:everything}):
 the first term on the right-hand side in Eq.~(\ref{eq:everything}) contains 
the residue of $A_e^{\mu\nu}$ at $d>4$, which itself contains -- among other contributions -- the pion decay constant $f_\pi$ in pure QCD. At this order in the low energy expansion, one may therefore factor out $f_\pi^2$ in the expression for the width~\cite{knecht}.
 The final result may be written \cite{knecht} in the form given by Marciano and Sirlin (Eq. (7a) of Ref.~\cite{marciano}),
\bea\label{eq:finalwidth}
\Gamma (\pi\to \ell\nu\, (\gamma))
&=&\frac{G_\mu^2|V_{ud}|^2f_\pi^2M_{\pi^+}m_\ell^2}{8\pi}(1-z_\ell)^2\times\nonumber\\
&&\left\{1+\frac{\alpha}{\pi}
\left[\log{\frac{M_Z^2}{m_\rho^2}}-\frac{3}{2}\log{\frac{m_\rho}{M_{\pi^+}}}+F(\sqrt{z_\ell})-C_1\right]\right\}\scs
\eea
with $z_\ell=m_\ell^2/M_{\pi^+}^2$. The function $F(x)$ is given in Eq.~(7b) of Ref.~\cite{marciano}, and the parameters $C_{2,3}$ introduced there do not occur at this order in the low 
energy expansion~\cite{knecht}. The constant $C_1$ can be expressed in 
terms of LECs and mass logarithms, see Eq.~(5.11) of Ref.~\cite{knecht} [its adaption to the $SU(2)\times SU(2)$ case considered here is straightforward]. It is now clear that, once $G_\mu, V_{ud}$ and $C_1$ are known, the pion decay constant is fixed 
through data on $\pi\to \ell\nu(\gamma)$. This is how the numerical result Eq.~(\ref{eq:valueoffpi}) was obtained by PDG [where higher order corrections 
in the width, as worked out in Ref.~\cite{cirigliano}, were taken into account as well.].

{\bf{Answer to question ii):}} 

Here arise two subtle points.

a) In factoring out $f_\pi$ in Eq.~(\ref{eq:finalwidth}), one makes use of 
the chiral expansion  of $f_\pi$ as given in Eq.~(\ref{eq:fpiwhatmass?}).
The choice of the mass $M$ in that expansion amounts to a
 convention: one may use either the  neutral or the charged pion mass.
 Scrutinizing the calculations performed in 
Refs.~\cite{knecht,cirigliano}, we find that the value Eq.~(\ref{eq:valueoffpi}) corresponds 
to the choice  $M=M_{\pi^0}\simeq135$ MeV [physical value of the neutral pion mass].  
Factoring 
 out $f_\pi$ evaluated at $M=M_{\pi^+}=139.57$ MeV [physical value of the charged pion mass] amounts 
to a renormalization of the constant $C_1$.
The formula (\ref{eq:fpiwhatmass?}) allows one to determine the difference between the two cases,
\bea\label{eq:valuefpimpi}
f_\pi(M_{\pi^+})&=&f_\pi(M_{\pi^0})+\frac{M_{\pi^+}^2-M_{\pi^0}^2}{8\pi^2f}\bigl(\bar l_4-1\bigr)+{\mathcal{O}}[p^4,(M^2_{\pi^+}-M^2_{\pi^0})^2]\scs\nonumber\\
\bar l_4&=&16\pi^2l_4^r-\ln{\frac{M_{\pi^+}^2}{\mu^2}}\fs
\eea
Using the value $\bar l_4=4.4$ \cite{pipinpb}, we find 
\be
f_\pi(M_{\pi^+})=f_\pi(M_{\pi^0})+0.4\, \mbox{MeV}\fs
\ee
The difference is thus quite significant -- about twice the 
uncertainty reported in the PDG-value Eq.~(\ref{eq:valueoffpi}). The induced change in $C_1$ is
\bea
C_1\to C_1+2.8\fs
\eea

b) Concerning the second point, we note that,
extracting a value for $f_\pi$ [defined in pure QCD, e.g. through Eq.~(\ref{eq:twopoint})] 
 {}from leptonic pion decays, where real and virtual photons are included, 
requires that one 
performs a splitting between strong and electromagnetic effects. This splitting is ambiguous~\cite{physrep} - 
the result depends on the procedure chosen. 
It is at this stage that the matching of the effective theory to the underlying 
theory matters. For a detailed analysis of this fact in a case which can be analyzed in a
 perturbative manner [linear sigma model, coupled to electromagnetism],
 we refer the reader to Ref.~\cite{gasscirus}. Here we note that,  in 
the language of the frameworks used in 
Refs.~\cite{moussallam,knecht,ananthanarayanmoussallam,descotesgenon,cirigliano}, 
the quantity $f_\pi$ depends on the scale of the underlying 
theory \cite{ananthanarayanmoussallam1,gasscirus}: $\mu_0\frac{df_\pi}{\mu_0}={\mathcal{O}}[e^2m_q]$. A different method to perform the matching 
consists in evaluating
 quantities in QCD with values of the parameters  relevant in QCD+QED at a 
scale $\mu_1$. In this case, strong quantities are scale independent, but do 
depend on the matching scale $\mu_1$.
  This scenario is discussed in detail in 
Refs.~\cite{gasscirus,rusetskycd09}, see also Refs.~\cite{leutwylercd09,flag}. For the pion decay constant in the chiral limit, the dependence on $\mu_1$ can be worked out in the framework of QCD, using the observation that $f$ is proportional to the renormalization group invariant scale of QCD, and applying the formula Eq.~(11.6) in Ref.~\cite{physrep}\footnote{We are indebted to H. Leutwyler for pointing this out to us.}. It turns out that, if $\mu_1$ is changed by a factor of 2, $f$ changes by a negligible amount of about 8 keV. The effect is so small, because the electromagnetic renormalization of the strong 
coupling constant $g$ starts out at two-loop order and is $\mathcal{O}(e^2g^3)$~\cite{physrep}- the one-loop 
contribution, which would be  of  order $e^2g$,  vanishes. In the linear sigma model, the scale 
dependence of $f$ is more than
 one order of magnitude larger~\cite{gasscirus}.

In our opinion, it would be very useful to perform the matching of the effective 
theory constructed   in Refs.~\cite{urech,neufeld96,knecht,cirigliano} to the underlying 
theory [the Standard Model] in this setting,
 which is used in  \cite{flag}.

\section{Effects from $m_d\neq m_u$}
\vspace{-2\parskip}
Finally, we comment on  $m_d-m_u$ effects in pure QCD, and note that the difference 
$m_d-m_u$ can occur only with even powers in $f_\pi$~\cite{bijnensprivate}. 
One-loop contributions in ChPT are linear in the quark masses, and thus cannot  contain 
isospin breaking terms, 
 whereas they do occur at and beyond two-loop order. Indeed, the latter
 have been evaluated in Ref.~\cite{kampfmoussallam} and found to be 
tiny, $f_{\pi^+}/f_{\pi^0}-1\simeq  0.7\times 10^{-4}$. Barring unexpectedly large higher order contributions,
  one concludes~\cite{kampfmoussallam} that $f_{\pi^+}\simeq f_{\pi^0}$ in pure QCD, to a very good approximation.  A numerical estimate of the contributions at order $e^2$ in 
$f_{\pi^+}/f_{\pi^0}-1$ can be found in  Ref.~\cite{steininger}.

\section{Summary}
\vspace{-2\parskip}
We have evaluated the correlator of two charged axial currents at one loop
 in ChPT, including virtual photons, for any value of the gauge fixing parameter $\xi$.
 As is seen {}from the result Eq.~(\ref{eq:A_limit1}), photon loops modify the holomorphic properties of the correlator  in a fundamental manner: the pole at the charged pion mass becomes a branch point, with strengths 
$\kappa_1,g_1$ that are gauge dependent. The residue  $\kappa_1^2$ furthermore depends on the scale of the underlying theory [QCD+QED], through the LEC $k_9^r$~\cite{moussallam,descotesgenon}.
For the evaluation of leptonic pion decays in the framework of the effective field theory framework set up in Refs.~\cite{urech,neufeld96,knecht,cirigliano}, one may use the standard LSZ formalism also at vanishing photon mass, provided that the graphs 
are worked out in $d$ space-time dimensions, and provided that the residue is evaluated at $d>4$~\cite{rad1,rad2}.  
We have discussed the manner in which the pion decay constant shows up
in the  final formula for the decay width, and have pointed out that the value reported
 in Eq.~(\ref{eq:valueoffpi})  corresponds to the case where the pion mass is identified with the neutral one.
 Evaluating $f_\pi$ at  the charged pion mass increases its value by about
0.4  MeV. Furthermore, we note  that  the value Eq.~(\ref{eq:valueoffpi}) 
is based on a matching procedure which differs {}from the one used in 
Ref.~\cite{flag}. In particular,  $f_\pi$ depends on the scale of the underlying theory.

\section*{Acknowledgments}
\vspace{-2\parskip}
We thank J.~Bijnens, G.~Colangelo, M.~Knecht, H.~Leutwyler, 
W.~J.~Marciano, B.~Moussallam, H.~Neufeld and  A.~Sirlin
for comments and/or for discussions. One of us (J.G.) thanks in particular 
A.~Rusetsky for many very enjoyable and informative discussions on the 
issue over the last years. Useful comments on the manuscript by G.~Colangelo, H.~Leutwyler, Ulf-G.~Mei{\ss}ner, H.~Neufeld and A.~Rusetsky are gratefully acknowledged.
An important part of this work was performed while GRSZ was visiting the ITP in 
Bern. He thanks the ITP for hospitality and for financial support. 
The work of GRSZ was supported by CAPES (Brazilian Agency).
 The Center for Research and Education in Fundamental Physics is
  supported by the ``Innovations- und Kooperationsprojekt C-13'' of
  the ``Schweizerische Universit\"atskonferenz SUK/CRUS''.
This work was  partially supported  by the Swiss
National Science Foundation,  by EU MRTN-CT-2006-035482
(FLAVIA{\it net}), and
by the Helmholtz Association through funds provided to the 
virtual institute 
``Spin and strong QCD'' (VH-VI-231). 

\section*{Appendix}
\vspace{-2\parskip}
\renewcommand{\theequation}{A\arabic{equation}}
\setcounter{equation}{0}

Here we prove Eq.~(\ref{eq:transitionampl}), which is true in the  absence 
of electromagnetic  interactions.  In this case, there are no virtual leptons in the corresponding Feynman diagrams, and one may evaluate the transition amplitude by 
considering the lepton current as a classical external field which we denote by $X_\mu$,
such that
\be
l_\mu=v_\mu-a_\mu-X_\mu\scs r_\mu=v_\mu+a_\mu\scs
\ee
and
\be
\bar{\mathcal L}_{eff} = \frac{F^2}{4}
\left\langle u_\mu u^\mu+\chi_+ \right\rangle 
\ee
is the pertinent leading order Lagrangian, which has the structure
\be
\bar{\mathcal L}_{eff} = {{\mathcal L}_{\pi\pi}}+\langle \ell_\mu L^\mu\rangle +\langle r_\mu R^\mu\rangle +{\mathcal O}(r^2,l^2,rl)\,;\,
{{\mathcal L}_{\pi\pi}}=\bar{\mathcal L}_{{eff}_{|l= r = 0}}\scs
\ee
with operator-valued currents $L_\mu,R_\mu$. Let $L^\mu\pm R^\mu=O_\pm^\mu$, with $O_\pm^\mu(\phi)=\pm O_\pm^\mu(-\phi).$ 
At $v_\mu=0$, the terms quadratic in the external fields in the $S$-operator are 
\be\label{eq:S-operator}
S=\int dx dyT e^{i\int {\mathcal L}_{\pi\pi}(z)dz}\langle a_\mu
O_-^\mu\rangle_x \langle (a_\rho+X_\rho)O_-^\rho\rangle_y\scs
\ee
up to contact terms, which do not contribute to the matrix element in question. The symbol $T$ denotes time ordering.  {}From Eq.~(\ref{eq:S-operator}), it is seen that the coefficient of
{\it axial current $\times$axial current} is identical to the coefficient of {\it axial current $\times$lepton current}. 
 It is furthermore  clear {}from the derivation 
that this statement remains true in the presence  of additional terms in the Lagrangian 
which contain lepton fields exclusively through their presence in  the left 
current $l_\mu$. In particular, the coefficients are still the same 
in the presence of strong  counterterms.
{}From these observations  and {}from 
Eqs.~(\ref{eq:LSZ},\ref{eq:twopoint0},\ref{eq:twopoint}) follows 
Eq.~(\ref{eq:transitionampl}). Eq.~(\ref{eq:classCI}) is true, because the above
statements also hold in the presence of 
 an external electromagnetic field and of the mass term 
$e^2F^4Z \left\langle u^\dagger Q u^2Qu^\dagger \right\rangle$.

\ed